\documentclass{iopart}
\usepackage{iopams}  
\newtheorem{theorem}{Theorem}[section]
\newtheorem{proposition}[theorem]{Proposition}
\newtheorem{lemma}[theorem]{Lemma}

\eqnobysec
\begin{document}
%%%%%%%%%%%%%%%%%%%%%%%%%%%%%%%%%%%%%%%%%%%%%%%%%%%
%%% title
%%%%%%%%%%%%%%%%%%%%%%%%%%%%%%%%%%%%%%%%%%%%%%%%%%%
\title{Hypergeometric solutions to the $q$-Painlev\'e equation of type $A_4^{(1)}$}
\author{Taro Hamamoto and Kenji Kajiwara}
\address{Graduate School of Mathematics, Kyushu University, 6-10-1 Hakozaki, Fukuoka 812-8581, Japan}
%%%%%%%%%%%%%%%%%%%%%%%%%%%%%%%%%%%%%%%%%%%%%%%%%%%
%%% Abstract
%%%%%%%%%%%%%%%%%%%%%%%%%%%%%%%%%%%%%%%%%%%%%%%%%%%
\begin{abstract}
We consider the $q$-Painlev\'e equation of type $A_4^{(1)}$ (a version
of $q$-Painlev\'e V equation) and construct a family of solutions
expressible in terms of certain basic hypergeometric series. We also present
the determinant formula for the solutions.
\end{abstract}
\ams{34M55, 39A13, 33D15, 33E17}
%\maketitle
\section{Introduction}
In this article we consider the $q$-difference equation
\begin{equation}
\begin{array}{c}
\left\{
\begin{array}{l}\medskip
 {\displaystyle \overline{g}g=\frac{qt}{b_2}~\frac{(f+b_3)(f+1)}{f+{\displaystyle \frac{1}{s}}},}\\
\medskip
{\displaystyle
\overline{f}f=\frac{1}{s}~\frac{\left(\overline{g}+{\displaystyle \frac{1}{b_2}}\right)
\left(\overline{g}+{\displaystyle \frac{1}{b_1b_2}}\right)}
{\overline{g}+{\displaystyle \frac{qt}{b_2}}},}
\end{array}\right.\\
{\displaystyle \overline{b}_i=b_i\ (i=1,2,3),\quad \overline{t}=qt,\quad s=\frac{1}{qb_1b_2b_3t},}
\end{array}\label{eq:qp5} 
\end{equation}
where $q$ is a constant and $\overline{\phantom{f}}$ denotes the
discrete time evolution. \eref{eq:qp5} can be also expressed as
\begin{equation}
\begin{array}{c}
\left\{
\begin{array}{l}\medskip
{\displaystyle \overline{y}y=\frac{{\displaystyle \left(x+\frac{a_1}{z}\right)\left(x+\frac{1}{a_1z}\right)}}{{\displaystyle 1+a_3x}}},\\
{\displaystyle x\underline{x}=\frac{{\displaystyle \left(y+\frac{a_2}{\rho}\right)\left(y+\frac{1}{a_2\rho}\right)}}
{{\displaystyle 1+\frac{y}{a_3}}}},
\end{array}\right.\\
{\displaystyle \overline{a}_i=a_i\ (i=1,2,3),\quad \overline{z}=qz,\quad z=q^{\frac{1}{2}}\rho,}
\end{array}\label{eq:qp5_xy}
\end{equation}
where the variables are related as
\begin{equation}
\fl
 b_1 = a_2^2,\quad b_2 = \frac{1}{q^{\frac{1}{2}} a_1a_2a_3^2},\quad
  b_3=a_1^2,\quad t=\frac{a_3}{q^{\frac{1}{2}}a_2}z,\quad f=a_1zx,\quad g=a_1a_3^2zy.
\end{equation}
\eref{eq:qp5_xy} was first derived and identified as one of the discrete
Painlev\'e equations with a continuous limit to the Painlev\'e V
equation in \cite{KTGR:asymmetric}.  Sakai has classified \eref{eq:qp5}
as the discrete dynamical system on the rational
surface of type $A_4^{(1)}$ which admits the symmetry of affine Weyl
group of type $A_4^{(1)}$ \cite{Sakai}. Geometrical structure of the
$\tau$ functions on the $A_4$ weight lattice has been investigated in
\cite{RGO:qp5} as well as various B\"acklund transformations. 
In this article, we denote \eref{eq:qp5} (or \eref{eq:qp5_xy})
as dP($A_4^{(1)}$) following the notation that was adopted in
\cite{MSY}. We also write \eref{eq:qp5} as
dP($A_4^{(1)}$)$[b_1,b_2,b_3]$ when it is necessary to specify the
values of parameters explicitly.

It is well-known that the Painlev\'e and discrete Painlev\'e equations
admit two classes of particular solutions; hypergeometric solutions and
algebraic solutions. In particular, the determinant formula for the
hypergeometric solutions play an important role in applications, for
example, to the area related to matrix integration, such as random
matrix theory
\cite{Adler-Moerbeke,Borodin-Boyarchenko,Borodin:dP,Chen-Feigin,Forrester:growth_and_P,%
Forrester-Witte:biorthogonal,Forrester-Witte:P6tau,Forrester-Witte:tau_to_random,%
Forrester-Witte:dP_and_orthogonal,Forrester-Witte:discrete,OKS:dP1,Tracy-Widom}.
The simplest hypergeometric solution to dP($A_4^{(1)}$) has been obtained in
\cite{KMNOY:hyper1,KMNOY:hyper2,RGTT:special_fn}.  The purpose of this
article is to construct hypergeometric solutions to dP($A_4^{(1)}$)
\eref{eq:qp5} and present the determinant formula.  In section
2, we construct the simplest hypergeometric solution through the Riccati
equation which is reduced from \eref{eq:qp5} by imposing a condition on
the parameters. By applying a B\"acklund transformation we construct
complex hypergeometric solutions and present the determinant formula in
section 3. We give the proof in section 4.
%%%%%%%%%%%%%%%%%%%%%%%%%%%%%%%%%%%%%%%%%%%%%%%%
% Section 2
%%%%%%%%%%%%%%%%%%%%%%%%%%%%%%%%%%%%%%%%%%%%%%%%
\section{Riccati solution}
We first recall the definition of the basic hypergeometric series\cite{Gasper-Rahman}
\begin{equation}
\fl
 {}_r\phi_{s}\left[\begin{array}{c}a_1,\ldots,a_r\\b_1,\ldots,b_s\end{array};~q,z \right]
=\sum_{n=0}^\infty
 \frac{(a_1,\ldots,a_r;q)_{n}}{(b_1,\ldots,b_s;q)_{n}(q;q)_{n}}
 \left[(-1)^n q^{\frac{n(n-1)}{2}}\right]^{1+s-r}z^n,
\end{equation}
where
\begin{equation}
 (a_1,\ldots,a_r;q)_n=\prod_{i=1}^r (a_i;q)_n,\quad (a;q)_n = \prod_{k=1}^{n}(1-aq^{k-1}).
\end{equation}
The simplest solution that is expressible in terms of the basic
hypergeometric function is constructed by looking for the special case
in which dP($A_4^{(1)}$) \eref{eq:qp5} is reduced to the Riccati equation.
In fact, imposing the condition on \eref{eq:qp5}
\begin{equation}
 b_2=1
\end{equation}
then it admits a specialization $1+f+g=0$ to yield the
Riccati equation
\begin{equation}
\fl
\overline{g}=-qt \frac{g+1-b_3}{g+1-qt b_1 b_3} \ , \
\overline{f}=-(1+\overline{g})=\frac{qt b_3 (1-b_1)+(qt-1)f}{qt b_1 b_3 +f}.\label{eq:riccati}
\end{equation}
Linearizing the Riccati equation \eref{eq:riccati} by the standard technique, one
obtains the following solution (see also \cite{RGTT:special_fn,KMNOY:hyper1,KMNOY:hyper2}):
\begin{proposition}\label{prop:riccati}
Let $\psi=\psi(t,b_1,b_3)$ a function satisfying 
\begin{eqnarray}
&&\psi(qt,b_1,b_3)=b_1  \psi(t,b_1,b_3)+(1-b_1)~\psi(qt,b_1/q,b_3),  \label{eq:re1}\\
&&b_3\psi(qt,b_1,b_3)=\psi(t,b_1,b_3)+(b_3 -1)~\psi(t,b_1,qb_3), \label{eq:re2}\\
&&qt b_1 b_3\psi(q t,b_1,b_3) =(qtb_1-1)~\psi(t,b_1,b_3)  + \psi(t,q b_1,b_3), \label{eq:re3}\\
&&qt\psi(qt,b_1,b_3) =(qt b_1-1)~\psi(t,b_1,b_3)  +\psi(qt,b_1,b_3/q).\label{eq:re4}
\end{eqnarray}
Then 
\begin{equation}\label{eq:pr}
f=qtb_{3}(1-b_{1})\frac{\psi(qt,b_{1}/q,qb_{3})}{\psi(t ,b_{1},qb_{3})},\ 
g=- \frac{\psi(t,b_{1},b_{3})}{\psi(t,b_{1},q b_{3})},
\end{equation}
gives a solution of dP($A_4^{(1)}$) \eref{eq:qp5} with $b_2 = 1$.
\end{proposition}
It should be remarked that several basic hypergeometric functions
satisfy the contiguous relations \eref{eq:re1}-\eref{eq:re4}\cite{Masuda:private}. For
example, we have
\begin{enumerate}
 \item 
\begin{equation}
\fl
\psi(t,b_{1},b_{3}) =
{}_2\phi_{1}\left[\begin{array}{c}1/b_{1},b_{3}\\0\end{array};~q,qt b_1 \right],
\label{eq:hyper1}
\end{equation}
 \item 
\begin{eqnarray}
\fl
 \psi(t,b_{1},b_{3}) &=&\frac{(qt,1/t,b_3;q)_\infty}{(qtb_1,b_1b_3;q)_\infty}
~{}_2\phi_{1}\left[\begin{array}{c}q/b_{3},0\\q/b_1b_3\end{array};~q,1/tb_1  \right]\label{eq:hyper2}\\
\fl
&=&
\frac{(qt,1/t,b_3;q)_\infty}{(qtb_1,b_1b_3;q)_\infty}
~{}_1\phi_{1}\left[\begin{array}{c}b_{3}/q\\b_1b_3/q\end{array};~1/q,1/qt  \right],\label{eq:hyper3}
\end{eqnarray}
 \item 
\begin{equation}
\fl
  \psi(t,b_{1},b_{3}) =\frac{(b_3t,q/b_3t;q)_\infty}{(qtb_1,qb_1,q/b_3;q)_\infty}
~{}_2\phi_{1}\left[\begin{array}{c}b_{3}/q,1/qb_1\\0\end{array};~1/q,tb_1 \right].\label{eq:hyper4}
\end{equation}
\end{enumerate}

In order to prove Proposition \ref{prop:riccati} we use the following Lemma:
\begin{lemma}\label{lemma:contiguity}
$\psi(t,b_1,b_3$) satisfy the contiguous relations
\begin{eqnarray}
&& qtb_3 \psi(qt,b_1,qb_3) = \psi(t,b_1,b_3) - (1-qtb_1b_3)~\psi(t,b_1,qb_3), \label{eq:re5}\\
&& tb_3 \psi(qt,b_1,b_3) = \psi(t/q,qb_1,b_3) - (1-t)~\psi(t,b_1,b_3).\label{eq:re6}
\end{eqnarray}
\end{lemma}
\noindent\textbf{Proof of Lemma \ref{lemma:contiguity}.} Eliminating $\psi(t,b_1,b_3)$ from
\eref{eq:re1} and \eref{eq:re3} we have
\begin{equation}
\fl
(qt b_1 b_3 -qt +\frac{1}{b_1})~\psi(qt, b_1 , b_3) 
+ (qt b_1-1)(\frac{1}{b_1}-1)\psi(qt,b_1/q ,b_3)=\psi(t,qb_1 , b_3).\label{eq:re7}
\end{equation}
Similarly, 
eliminating $\psi(qt,b_1/q ,b_3)$ from \eref{eq:re1} and \eref{eq:re3}$_{b_1 \rightarrow b_1/q}$
we obtain
\begin{equation}
\fl
tb_1 b_3 \psi(qt,b_1,b_3) - (t {b_1}^{2} b_3 +1- b_1 )~\psi(t,b_1,b_3)
=(1- b_1 )(t b_1 -1)\psi(t,b_1/q ,b_3).\label{eq:re8}
\end{equation}
Then eliminating $\psi(qt,b_1/q,b_3)$ from \eref{eq:re7} and \eref{eq:re8}$_{t \rightarrow qt}$
we get 
\begin{equation*}
(1-qt)~\psi(qt, b_1, b_3)+qtb_3~\psi(q^{2}t , b_1 ,b_3 )=\psi(t,qb_1,b_3),
\end{equation*}
which is nothing but \eref{eq:re6}$_{t \rightarrow qt }$. Similarly, \eref{eq:re5}
can be derived by eliminating $\psi(qt,b_1,b_3)$ from \eref{eq:re2} and 
\eref{eq:re4}$_{b_3 \rightarrow qb_3}$.
$\square$\\

Proposition \ref{prop:riccati} follows immediately from Lemma
\ref{lemma:contiguity}.  In fact, dividing \eref{eq:re2} by
\eref{eq:re5} we have
\begin{displaymath}
\fl
\frac{1}{qt}\frac{\psi(qt,b_1,b_3)}{\psi(qt,b_1,qb_3)}
=- \frac{\overline{g}}{qt}=
\frac{\psi(t,b_1,b_3)+ (b_3 -1)~\psi(t,b_1,qb_3)}{\psi(t,b_1,b_3)-(1-qt b_1b_3)~\psi(t,b_1,qb_3)}
=\frac{-g+(b_3 -1)}{-g-(1-qt b_1 b_3)},
\end{displaymath}
which is the first equation of \eref{eq:riccati}. The second equation of 
\eref{eq:riccati} can be derived in similar manner by dividing 
\eref{eq:re6}$_{t\rightarrow qt,b_1 \rightarrow b_1/q,b_3 \rightarrow qb_3}$
by \eref{eq:re1}$_{b_3 \rightarrow qb_3}$ . $\square$
%%%%%%%%%%%%%%%%%%%%%%%
% section 3
%%%%%%%%%%%%%%%%%%%%%%%
\section{Determinant formula and bilinear equations}
%%%%%%%%%%%%%%%%%%%%%%%%%%%%%%%%%%%%%%%%%%%%%%%%
% section 3.1
%%%%%%%%%%%%%%%%%%%%%%%%%%%%%%%%%%%%%%%%%%%%%%%%
\subsection{B\"acklund transformations}
Sakai constructed the following transformations for the homogeneous
variables $x$, $y$, $z$ of $\mathbb{P}^2$ and the parameters $b_i$
($i=0,1,2,3,4$) on the $A_4^{(1)}$ type ({\it Mul.5})
surface\cite{Sakai}\footnote{Actions of these transformations are slightly modified
from the original formula to be subtraction-free.}:
\begin{eqnarray}
\fl&& \sigma:\ \left(\begin{array}{c}b_1\ b_2\ b_3\\b_4\ b_0\end{array};~x:y:z\right) \mapsto \nonumber\\
\fl&&
\left(\begin{array}{c}b_3\ b_4\ b_0\\b_1\ b_2\end{array};~b_4xy(z+x): b_2z(x+y+z)(x+b_4y+z):x(x+z)^2\right),
\label{eq:weyl_sigma}
\\
\fl&& \sigma':\ \left(\begin{array}{c}b_1\ b_2\ b_3\\b_4\ b_0\end{array};~x:y:z\right)\mapsto \nonumber\\
\fl&& 
\left(\begin{array}{c}\frac{1}{b_1}\ \frac{1}{b_0}\ \frac{1}{b_4}\\\frac{1}{b_3}\ \frac{1}{b_2}\end{array};
~b_2z(x+z)(x+y+z): y((z+x)(b_0x+b_2z)+b_2yz):b_0x(x+z)^2\right),
\label{eq:weyl_sigma'}\\
\fl&& w_3:\ \left(\begin{array}{c}b_1\ b_2\ b_3\\b_4\ b_0\end{array};~x:y:z\right)\mapsto \nonumber\\
\fl&&
\left(\begin{array}{c}b_1\ b_2b_3\ \frac{1}{b_3}\\b_3b_4\ b_0\end{array};
~b_3x(b_3x+y+b_3z): y(b_3x+y+z):b_3z(x+y+z)\right),
\label{eq:weyl_w3}
\\
\fl&& w_1:\ \left(\begin{array}{c}b_1\ b_2\ b_3\\b_4\ b_0\end{array};~x:y:z\right)\mapsto 
\left(\begin{array}{c}\frac{1}{b_1}\ b_1b_2\ b_3\\b_4\      b_1b_0\end{array};~x:y:z\right),
\label{eq:weyl_w1}\\
\fl&& w_2:\ \left(\begin{array}{c}b_1\ b_2\ b_3\\b_4\ b_0\end{array};~x:y:z\right)\mapsto 
\left(\begin{array}{c}b_1b_2\ \frac{1}{b_2}\ b_2b_3\\b_4\ b_0\end{array};~x:b_2y:b_2z\right),
\label{eq:weyl_w2}\\
\fl&& w_4:\ \left(\begin{array}{c}b_1\ b_2\ b_3\\b_4\ b_0\end{array};~x:y:z\right)\mapsto 
\left(\begin{array}{c}b_1\ b_2\ b_3b_4\\\frac{1}{b_4}\     b_4b_0\end{array};~x:b_4y:z\right),
\label{eq:weyl_w4}
\\
\fl && w_0 = \sigma^2\circ w_1\circ \sigma^3.\label{eq:weyl_w0}
\end{eqnarray}
Introducing the variables $f$ and $g$ by
\begin{equation}
 f = \frac{y}{z+x},\quad g=\frac{z(x+y+z)}{x(z+x)},
\end{equation}
then \eref{eq:weyl_sigma} - \eref{eq:weyl_w0} can be rewritten as
%%%%%%%%%%%%%%%%%%%%%%%%%%%%%%%%%%%%%%%%%%%%%%%%%%%%%
% Note!!
% w_i in Hamamoto's thesis corresponds to our w_{i-1}
%
%%%%%%%%%%%%%%%%%%%%%%%%%%%%%%%%%%%%%%%%%%%%%%%%%%%%%
\begin{equation}\label{eq:weyl_fg}
\fl
\begin{array}{l}
\smallskip
\sigma: (b_0,b_1,b_2,b_3,b_4,f,g) \mapsto (b_2,b_3,b_4,b_0,b_1, b_2 g,\frac{1 + b_2 g}{b_4f}),\\
\smallskip
\sigma': (b_0,b_1,b_2,b_3,b_4,g) \mapsto 
(\frac{1}{b_1},\frac{1}{b_0},\frac{1}{b_4},\frac{1}{b_3},\frac{1}{b_2},\frac{b_0(1 + g)}{b_2f}),\\
\smallskip
w_0: (b_0,b_1,b_4,f,g) \mapsto
(\frac{1}{b_0}, b_0 b_1,b_0 b_4, \frac{f(b_0 + b_2 g)}{b_0 (1+b_2 g)},\frac{g}{b_0}),
\\
\smallskip
w_1: (b_0,b_1,b_2) \mapsto (b_0 b_1,\frac{1}{b_1},b_1 b_2),\\
\smallskip
w_2: (b_1,b_2,b_3,f,g) \mapsto  
(b_1 b_2,\frac{1}{b_2},b_2 b_3, b_2f\frac{1+f+g}{1+f+b_2g},b_2g\frac{1+b_2f+b_2g}{1+f+b_2 g}),\\
\smallskip
w_3: (b_2,b_3,b_4,f,g) \mapsto (b_2 b_3,\frac{1}{b_3},b_3 b_4,\frac{f}{b_3},\frac{g}{b_3}),\\
w_4: (b_0,b_3,b_4,f,g) \mapsto (b_0 b_4,b_3 b_4,\frac{1}{b_4}, b_4f ,\frac{g(1 + b_4f)}{1+f}),
   \end{array}
\end{equation}
respectively, where the abbreviated variables are invariant with respect to the
transformation. It can be shown by direct calculation that these
transformations satisfy the fundamental relation of the (extended)
affine Weyl group $\widetilde{W}(A_4^{(1)})$:
\begin{equation}
\fl
\begin{array}{l}
\smallskip
{\displaystyle w^{2}_{i}=1,\ (w_{i} w_{i\pm 1})^{3}=1,\
(w_{i}w_{j})^{2}=1 \ (j \not\equiv i,i\pm 1),\  \sigma^5=1,\ \sigma'^2=1,}\\
{\displaystyle \sigma w_{i}=w_{i+2} \sigma,\ \sigma' w_0=w_2 \sigma',\ 
\sigma' w_3=w_4\sigma',\ \sigma' w_1=w_1\sigma',\ i \in \mathbb{Z}/5\mathbb{Z}}.
\end{array}
\end{equation}
We note that $q=1/(b_0b_1b_2b_3b_4)$ is invariant with respect to the
Weyl group actions. The translation $T_0=w_4w_3w_2w_1\sigma^{3}$ 
acts on $b_i$ as
\begin{equation}
 T_0:(b_0,b_1,b_2,b_3,b_4)\mapsto (qb_0,b_1,b_2,b_3,b_4/q),
\end{equation}
and the action on $f$ and $g$ is nothing but dP($A_4^{(1)}$)\eref{eq:qp5} for $t=b_0$ and $s=b_4$.
If we define the translations $T_i$ ($i=1,2,3,4$) by
\begin{equation}
 T_1 = \sigma^3 T_0 \sigma^2,\quad T_2=\sigma T_0 \sigma^4,\quad
T_3=\sigma^4 T_0 \sigma,\quad T_4=\sigma^2 T_0 \sigma^3,
\end{equation}
then actions of $T_i$ ($i=1,2,3,4$) on the parameters are given by
\begin{equation}
 \begin{array}{l}
{\displaystyle T_1:(b_0,b_1,b_2,b_3,b_4)\mapsto (b_0/q,qb_1,b_2,b_3,b_4),}\\
{\displaystyle T_2:(b_0,b_1,b_2,b_3,b_4)\mapsto (b_0,b_1/q,qb_2,b_3,b_4),  }\\
{\displaystyle T_3:(b_0,b_1,b_2,b_3,b_4)\mapsto (b_0,b_1,b_2/q,qb_3,b_4),}\\
{\displaystyle T_4:(b_0,b_1,b_2,b_3,b_4)\mapsto (b_0,b_1,b_2,b_3/q,qb_4).  }
 \end{array}
\end{equation}
and one can directly verify that $T_iT_j=T_jT_i$ ($i,j=0,1,2,3$, $i\neq
j$) and $T_0T_1T_2T_3T_4=1$.  Therefore if we regard $T_0$ as the
discrete time evolution, $T_i$ ($i=1,2,3,4$) can be regarded as the
B\"acklund transformations. 
%
%%%%%%%%%%%%%%%%%%%%%%%%%%%%%%%%%%%%%%%%%%%%%%%%
% section 3.2
%%%%%%%%%%%%%%%%%%%%%%%%%%%%%%%%%%%%%%%%%%%%%%%%
\subsection{Determinant formula}
Let us apply the B\"acklund transformation $T_2$ on the Riccati solution
obtained in Proposition \ref{prop:riccati}. Applying $T_2$ $N$ times
yields the solution for dP($A_4^{(1)}$)$[q^{-N}b_1,q^N,b_3]$,
which is expressed as rational function in $\psi$. However the
denominator and numerator can be factorized into two factors,
respectively, and each factor admits determinant formula. More
precisely, we have the following formula, which is the main result of
this article:
\begin{theorem}\label{thm:det}
Let $\tau_N(t,b_1,b_3)$ ($N\in\mathbb{Z}$) be 
\begin{equation}
\fl \tau_N(t,b_1,b_3)=\left\{
\begin{array}{cl}
\medskip
{\displaystyle \det(\psi(t,q^{-j+1}b_1,q^{i-1}b_3))_{i,j=1,\ldots,N}} 
& (N> 0),\\
\medskip
{\displaystyle 1} & (N=0),\\
{\displaystyle \det(\psi(t,q^{j-1}b_1,q^{-i+1}b_3))_{i,j=1,\ldots,M}}
& (N=-M<0),
\end{array}
\right.\label{eq:tau}
\end{equation}
Then 
\begin{equation}\label{eq:fg}
\fl
\begin{array}{l}
\medskip
f=
\left\{
\begin{array}{cl}
\medskip
{\displaystyle 
q^{N+1}t b_{3}(1 - q^{-N}b_1)~
\frac{\tau_{N}(t,b_1,qb_3)~\tau_{N+1}(qt,b_1/q,qb_3)}{\tau_{N}(qt,b_1/q,qb_3)~\tau_{N+1}(t,b_1,qb_3)}
} & (N\geq 0),\\
{\displaystyle -\frac{\tau_N(t,qb_1,b_3)~\tau_{N+1}(qt,b_1,b_3)}{\tau_N(qt,b_1,b_3)~\tau_{N+1}(t,qb_1,b_3)}} & (N<0),
\end{array}
\right.\\
g=
\left\{
\begin{array}{cl}
\medskip
{\displaystyle 
-\frac{\tau_{N}(t,b_1/q,q^2b_3)~\tau_{N+1}(t,b_1,b_3)}{\tau_{N}(t,b_1/q,qb_3)~\tau_{N+1}(t,b_1,qb_3)}
} & (N\geq 0),\\
{\displaystyle
 qt(b_3-1)\frac{\tau_{N}(t,b_1,qb_3)~\tau_{N+1}(t,qb_1,b_3/q)}{\tau_N(t,b_1,b_3)~\tau_{N+1}(t,qb_1,b_3)}}
 & (N<0),
\end{array}\right.
\end{array}
\end{equation}
satisfy dP($A_4^{(1)}$)$[q^{-N}b_1,q^N,b_3]$.
\end{theorem}
We introduce a notation for simplicity
\begin{equation}
 \tau_N(t,q^mb_1,q^nb_3)=\tau_N^{m,n}(t).
\end{equation}
Theorem \ref{thm:det} for $N\geq 0$ is a direct consequence of the following Proposition:
\begin{proposition}\label{prop:bl}
For $N\geq 0$, $\tau_N^{m,n}(t)$ satisfy the following bilinear difference equations.
\begin{eqnarray}
\fl
&&(1-q^{-N}b_1)\tau^{0,1}_{N}(t)~\tau^{-1,1}_{N+1}(qt)
+ q^{-N}b_1 \tau^{-1,1}_{N}(qt)~\tau^{0,1}_{N+1}{(t)}\nonumber\\
\fl
&&\qquad
- q^{-N}\tau^{-1,1}_{N}(t)~\tau^{0,1}_{N+1}(qt)=0, \label{eq:bl1} \\
\fl
&&qtb_3(1-q^{-N}b_1)\tau^{0,1}_{N}(t)~\tau^{-1,1}_{N+1}(qt)
+  q^{-N}\tau^{-1,1}_{N}(qt)~\tau^{0,1}_{N+1}(t)\nonumber\\
\fl
&&\qquad
-  \tau^{-1,2}_{N}(qt)~\tau^{0,0}_{N+1}(t) =0, \label{eq:bl2} \\
\fl
&&qt (1 - q^{-N}b_1)\tau^{0,1}_{N}(t)~\tau^{-1,1}_{N+1}(qt)
+q^{-N}\tau^{-1,1}_{N}(qt)~\tau^{0,1}_{N+1}(t)\nonumber\\
\fl
&&\qquad
-q^{-N}\tau^{-1,2}_{N}(t)~\tau^{0,0}_{N+1}(qt) =0, \label{eq:bl3} \\
\fl
&&q^{-N}\tau^{-1,1}_{N}(t)\tau^{0,1}_{N+1}(t)- q^{-N}b_1\tau^{-1,2}_{N}(t)\tau^{0,0}_{N+1}(t)\nonumber\\
\fl
&&\qquad 
- (1-q^{-N}b_1)(1- q^{-N}t b_1)\tau^{0,1}_{N}(t)\tau^{-1,1}_{N+1}(t)=0, \label{eq:bl4} \\
\fl
&&q^{-N}\tau^{-1,1}_{N}(t)~\tau^{0,1}_{N+1}(t)
- \tau^{-1,2}_{N}(t)~\tau^{0,0}_{N+1}(t)\nonumber\\
\fl
&&\qquad
+ q^{N+1}tb_3(1-q^{-N}b_1)~\tau^{0,1}_{N}(t/q) ~\tau^{-1,1}_{N+1}(qt)=0,  \label{eq:bl5} \\
\fl
&&q^{-N}t ~\tau^{-1,1}_{N}(t)~\tau^{0,1}_{N+1}(t)
- \tau^{-1,2}_{N}(t)~\tau^{0,0}_{N+1}(t)\nonumber\\
\fl
&&\qquad
+ (1 - q^{-N}t b_1)~\tau^{-1,1}_{N}(qt)~\tau^{0,1}_{N+1}(t/q)=0.\label{eq:bl6} 
\end{eqnarray}
\end{proposition}
In fact, Theorem \ref{thm:det} for $N\geq 0$ can be derived from Proposition
\ref{prop:bl} as follows. We have from \eref{eq:bl1} by using
\eref{eq:fg} 
\begin{equation}
% (1-q^{-N}b_1)\frac{\tau^{0,1}_{N}(t)~\tau^{-1,1}_{N+1}(qt)}{\tau^{-1,1}_{N}(qt)~\tau^{0,1}_{N+1}{(t)}}
f
+ \frac{1}{s}
= qtb_3\frac{\tau^{-1,1}_{N}(t)~\tau^{0,1}_{N+1}(qt)}{\tau^{-1,1}_{N}(qt)~\tau^{0,1}_{N+1}{(t)}}.
\end{equation}
We also have from \eref{eq:bl2} and \eref{eq:bl3}
\begin{eqnarray}
 &&
f + 1 = q^N\frac{\tau^{-1,2}_{N}(qt)~\tau^{0,0}_{N+1}(t)}{\tau^{-1,1}_{N}(qt)~\tau^{0,1}_{N+1}(t)},\\
&&
f + b_3
= b_3\frac{\tau^{-1,2}_{N}(t)~\tau^{0,0}_{N+1}(qt)}{\tau^{-1,1}_{N}(qt)~\tau^{0,1}_{N+1}(t)},
\end{eqnarray}
respectively. Therefore we obtain
\begin{equation}
\fl
 \frac{(f+1)(f+b_3)}{f+\frac{1}{s}} 
=
\frac{q^{N-1}}{t}~
\frac{\tau^{-1,2}_{N}(qt)\tau^{0,0}_{N+1}(qt)\tau^{0,0}_{N+1}(t)\tau^{-1,2}_{N}(t)}
{\tau^{-1,1}_{N}(qt)\tau^{0,1}_{N+1}(qt)\tau^{0,1}_{N+1}(t)\tau^{-1,1}_{N}(t)}
=\frac{q^{N-1}}{t}~ \overline{g}g,
\end{equation}
which is the first equation of \eref{eq:qp5}. Similarly, from
\eref{eq:bl4}$_{t\to qt}$, \eref{eq:bl5}$_{t\to qt}$ and
\eref{eq:bl6}$_{t\to qt}$ we get
\begin{eqnarray}
\fl &&1+b_1 \overline{g}
=q^{N}(1 - q^{-N}b_1)(1 - q^{-N+1}tb_1)
\frac{\tau^{0,1}_N(qt)\tau^{-1,1}_{N+1}(qt)}{\tau^{-1,1}_{N}(qt)\tau^{0,1}_{N+1}(qt)},\\
\fl&& 1+ q^{N} \overline{g}=-q^{2N+2} tb_3 (1- q^{-N} b_1 )
\frac{\tau^{0,1}_{N}(t)\tau^{-1,1}_{N+1}(q^{2}t)}{\tau^{-1,1}_{N}(qt)\tau^{0,1}_{N+1}(qt)}, \\
\fl&& qt + q^N \overline{g} = - q^{N} (1 - q^{-N+1} t b_1  )
\frac{\tau^{-1,1}_{N}(q^{2}t)\tau^{0,1}_{N+1}(t)}{\tau^{-1,1}_{N}(qt) ~\tau^{0,1}_{N+1}(qt)},
\end{eqnarray}
respectively. Then we have
\begin{equation}
 qtb_3\frac{(1+b_1 \overline{g})(1+ q^{N} \overline{g})}{qt + q^N \overline{g}}
=\overline{f}f,
\end{equation}
which is the second equation of \eref{eq:qp5}. Therefore we have
verified that Theorem \ref{thm:det} for $N\geq 0$ follows from the
bilinear equations \eref{eq:bl1} - \eref{eq:bl6} in Proposition
\ref{prop:bl}. We omit the proof for the case of $N<0$ since it can be
proved in similar manner.

%%%%%%%%%%%%%%%%%%%%%%%%%%%%%%%%%%%%%%%%%%%%%%%%
% section 4
%%%%%%%%%%%%%%%%%%%%%%%%%%%%%%%%%%%%%%%%%%%%%%%%
\section{Proof of Proposition \ref{prop:bl}} \label{sec:proof}
The bilinear equations \eref{eq:bl1} - \eref{eq:bl6} can be reduced to
the Pl\"ucker relations which are quadratic identities among the
determinants whose columns are properly shifted.  This can be done by
constructing ``difference formulae'' that relate the ``shifted''
determinants with $\tau_N^{m,n}(t)$ by using the contiguous relations of
$\psi$. This technique has been developed in \cite{OHTI:dKP,OKMS:RT} and
applied to various discrete Painlev\'e
equations\cite{HKW:qp2,K:qp3-2,KK:qp3-1,KNY:qp4,KOS:dP3,KOSGR:dP2,KYO:dP2,NSKGR:alt-dP2,OKS:dP1,Sakai:qp6_sol}.
In this section, we prove the bilinear equation \eref{eq:bl1}
as an example.  Since other bilinear equations
\eref{eq:bl2}-\eref{eq:bl6} can be proved in similar manner, we leave
the details in the appendix.

We first introduce the following notation:
\begin{eqnarray}
\fl
\tau^{m,n}_{N}(t)&=&
\left|\begin{array}{cccc}
 \psi(t,q^{m}b_{1},q^{n}b_{3}) & \psi(t,q^{m-1}b_{1},q^{n} b_{3}) &\cdots &  \psi(t,q^{m-N+1}b_{1},q^{n}b_{3}) \\
 \psi(t,q^{m}b_{1},q^{n+1}b_{3}) & \psi(t,q^{m-1}b_{1},q^{n+1} b_{3})& \cdots & \psi(t,q^{m-N+1}b_{1},q^{n+1}b_{3})  \\
     \vdots & \vdots  & \cdots & \vdots  \\
 \psi(t,q^{m}b_{1},q^{n+N-1}b_{3}) & \psi(t,q^{m-1}b_{1},q^{n+N-1}b_{3}) & \cdots &  \psi(t,q^{m-N+1}b_{1},q^{n+N-1}b_{3})
\end{array}\right| \nonumber \\
\fl 
&=&
\left|
\begin{array}{cccc}
\Psi_{m,n}(t)&\Psi_{m-1,n}(t)& \cdots&\Psi_{m-N+1,n}(t)
\end{array}
\right|, \label{eq:det1}
\end{eqnarray}
where $\Psi_{m,n}(t)$ denotes a column vector
\begin{equation}
\Psi_{m,n}(t)=
\left(\begin{array}{c}
\psi(t,q^mb_1,q^nb_3)\\
\psi(t,q^mb_1,q^{n+1}b_3)\\
\vdots\\
\psi(t,q^mb_1,q^{n+N-1}b_3)
\end{array}\right).
\end{equation}
Here we note that the height of the column vector is $N$, but we use the
same symbol for the column vector with different height. 
Then we have the following difference formula:
\begin{lemma}\label{lem:diff_formula}
\begin{eqnarray}
\fl
&&
\left|\begin{array}{cccc}
\Psi_{m,n}(t) &\Psi_{m,n}{(t/q)} & \cdots &  
\Psi_{m-N+2,n}{(t/q)}
\end{array}\right|\nonumber\\
\fl&&\hskip60pt
=\frac{\prod\limits^{N-2}_{k=0} (q^{m-k}b_1  -1)}
{q^{\frac{(N-1)(2m-N+2)}{2}}b_1^{N-1}}~\tau^{m,n}_{N}{(t)},\label{eq:cp-1}\\
\fl
&&
\left|\begin{array}{ccccc}
\Psi_{m,n}{(t/q)} &\Psi_{m-1,n}{(t)} &
\Psi_{m-1,n}{(t/q)}&\cdots &\Psi_{m-N+2,n}{(t/q)}
\end{array}\right| \nonumber\\
\fl&&\hskip60pt
=\frac{\prod\limits^{N-2}_{k=1} (q^{m-k}b_1-1)}{q^{\frac{(N-1)(2m-N+2)}{2}}b_1^{N-1}}
~\tau^{m,n}_{N}{(t)}.\label{eq:cp-2}
\end{eqnarray}
\end{lemma}
\noindent\textbf{Proof.}
Using the contiguous relation \eref{eq:re1} on the $N$-th column of the
determinant in \eref{eq:det1}, we have
\begin{eqnarray*}
\fl
\tau^{m,n}_{N}(t)&=&
\left|\begin{array}{ccccc}
\Psi_{m,n}(t)&\Psi_{m-1,n}(t)& \cdots &
\Psi_{m-N+2,n}(t) &\Psi_{m-N+1,n}(t)
 \end{array}\right|\\
\fl
&=&\left|\begin{array}{cccc}
\Psi_{m,n}(t)& \cdots & \Psi_{m-N+2,n}(t)
& \frac{-\Psi_{m-N+2,n}(t)+q^{m-N+2} b_1  
\Psi_{m-N+2,n}(t/q)}
{q^{m-N+2}b_1 -1}
 \end{array}\right|\\
\fl
&=&\frac{q^{m-N+2} b_1}{q^{m-N+2}b_1 -1}~
\left|\begin{array}{cccc}
\Psi_{m,n}(t)& \cdots &
\Psi_{m-N+2,n}(t)
& \Psi_{m-N+2,n}(t/q)
\end{array}\right|.
\end{eqnarray*}
Applying this procedure from the $N$-th column to the second column we obtain
\begin{equation}
\fl \tau_N^{m,n}(t)
=\frac{q^{\frac{(N-1)(2m-N+2)}{2}}b_1^{N-1}}{ {{\displaystyle
\prod^{N-2}_{k=0}}} (q^{m-k}b_1 -1)}~
\left|\begin{array}{cccc}
\Psi_{m,n}(t)&\Psi_{m,n}(t/q)& \cdots &   
\Psi_{m-N+2,n}(t/q)
\end{array}\right|,\label{eq:p-1}
\end{equation}
which is nothing but \eref{eq:cp-1}. At the stage where the above
procedure has been applied up to the third column, we have by using \eref{eq:re1}
on the first column
\begin{eqnarray}
\fl
&&\tau^{m,n}_{N}{(t)}\nonumber\\
\fl
&=&
\frac{q^{\frac{(N-2)(2m-N+1)}{2}}b_1^{N-2}}{\prod\limits_{k=1}^{N-2}(q^{m-k}b_1 -1)}
\left|\begin{array}{ccccc}
\Psi_{m,n}(t)&\Psi_{m-1,n}(t)
& \Psi_{m-1,n}(t/q)
& \cdots  & \Psi_{m-N+2,n}(t/q)
\end{array}\right|\nonumber\\
\fl
&=&\frac{q^{\frac{(N-2)(2m-N+1)}{2}}b_1^{N-2}}{\prod\limits_{k=1}^{N-2}(q^{m-k}b_1-1)}\nonumber\\
\fl
&\times&\left|\begin{array}{ccccc}
q^{m}b_1 \Psi_{m,n}(t/q) -(q^{m}b_1-1)\Psi_{m-1,n}(t)
&  \Psi_{m-1,n}(t)
&  \Psi_{m-1,n}(t/q)
&  \cdots  
& \Psi_{m-N+2,n}(t/q)
\end{array}\right|\nonumber\\
\fl
&=&\frac{q^{\frac{(N-1)(2m-N+2)}{2}}b_1^{N-1}}
{\prod\limits^{N-2}_{k=1}(q^{m-k}b_1 -1)}\nonumber\\
\fl
&\times&
\left|\begin{array}{ccccc}
\Psi_{m,n}{(t/q)} &\Psi_{m-1,n}{(t)} &
\Psi_{m-1,n}{(t/q)}&\cdots &\Psi_{m-N+2,n}{(t/q)}
\end{array}\right|,\label{eq:p-2}
\end{eqnarray}
which is \eref{eq:cp-2}. This completes the proof. $\square$\\

Now consider the Pl\"ucker relation
\begin{eqnarray}
\fl
0&=&
\left|\begin{array}{ccccc}
\Psi_{m+1,n}(t/q) & \Psi_{m,n}(t) &
\Psi_{m,n}(t/q) &\cdots&\Psi_{m-N+3,n}{(t/q)}\end{array}\right|\nonumber\\
\fl
&& \times  
\left|\begin{array}{ccccc}
\Psi_{m,n}{(t/q)} &\cdots&\Psi_{m-N+3,n}(t/q) &
\Psi_{m-N+2,n}(t/q) &\phi \end{array}\right|\nonumber\\
\fl
&-&
\left|\begin{array}{ccccc}
\Psi_{m+1,n}(t/q) &\Psi_{m,n}(t/q)  &\cdots&\Psi_{m-N+3,n}(t/q) &
\Psi_{m-N+2,n}(t/q) \end{array}\right|\nonumber\\
\fl
&&\times 
\left|\begin{array}{ccccc}
\Psi_{m,n}(t) &\Psi_{m,n}(t/q) &\cdots&
\Psi_{m-N+3,n}(t/q) &\phi \end{array}\right|\nonumber\\
\fl
&+&
\left|\begin{array}{ccccc}
 \Psi_{m+1,n}(t/q) &\Psi_{m,n}(t/q) &
\cdots&\Psi_{m-N+3,n}(t/q) &\phi\end{array}\right|\nonumber\\
\fl
&&\times
\left|\begin{array}{ccccc}
\Psi_{m,n}(t)&\Psi_{m,n}(t/q) &\cdots
 &\Psi_{m-N+3,n}(t/q) &\Psi_{m-N+2,n}(t/q)\end{array}\right|, \label{eq:Plucker}
\end{eqnarray}
where $\phi$ is the column vector
\begin{equation}
 \phi = \left(\begin{array}{c}0 \\\vdots \\ 0\\ 1 \end{array}\right).
\end{equation}
Applying Lemma \ref{lem:diff_formula} to \eref{eq:Plucker} we have
\begin{eqnarray*}
%\fl
&&\tau^{m+1,n}_{N}(t) ~\tau^{m,n}_{N-1}(t/q) - q^{m+1} b_1~\tau^{m+1,n}_{N}(t/q)~\tau^{m,n}_{N-1}(t)
\nonumber\\
%\fl
&&\hskip60pt + q^{N-1}(1 - q^{m-N+2} b_1)~\tau^{m+1,n}_{N-1}(t/q) ~\tau^{m,n}_{N}(t)=0.
\end{eqnarray*}
Putting $N \to N+1$, $t \to qt$, $m=-1$ and $n=1$ we obtain
\eref{eq:bl1}. $\square$

%%%%%%%%%%%%%%%%%%%%%%%%%%%%%%%%%%%%%%%%%%%%%%%%
% acknowledgments
%%%%%%%%%%%%%%%%%%%%%%%%%%%%%%%%%%%%%%%%%%%%%%%%
\ack The authors would like to thank T. Masuda, M. Noumi,
Y. Ohta, N.S. Witte and Y. Yamada for fruitful discussions. This work
has been supported by the JSPS Grant-in-Aid for Scientific Research
(B)15340057 and (A)16204007. The authors also acknowledge the support by
the 21st Century COE program at the Faculty of Mathematics, Kyushu
University.

%%%%%%%%%%%%%%%%%%%%%%%%%%%%%%%%%%%%%%%%%%%%%%%%
% appendix
%%%%%%%%%%%%%%%%%%%%%%%%%%%%%%%%%%%%%%%%%%%%%%%%
\appendix
\setcounter{section}{1}
\section*{Appendix: Proof of bilinear equations}
In this appendix we prove the bilinear equations \eref{eq:bl2}-\eref{eq:bl6}.
We first note that $\tau_N^{m,n}(t)$ admits various determinantal expressions,
which play an important role in proving the bilinear
equations. Taking the transpose of the right hand side of \eref{eq:det1}, we have
\begin{eqnarray}
\tau^{m,n}_{N}(t)&=&
\left|\begin{array}{cccc}
\widetilde\Psi_{m,n}(t)&\widetilde\Psi_{m,n+1}(t) & \cdots &\widetilde\Psi_{m,n+N-1}(t)
\end{array}\right|, \label{eq:det2}
\end{eqnarray}
where
\begin{equation}
\widetilde\Psi_{m,n}(t)=
\left(\begin{array}{c}
\psi(t,q^mb_1,q^nb_3)\\
\psi(t,q^{m-1}b_1,q^nb_3)\\
\vdots\\
\psi(t,q^{m-N+1}b_1,q^nb_3)
\end{array}\right).
\end{equation}
It is also possible to express $\tau_N^{m,n}(t)$ by the determinants with
different structure of shifts. 
\bigskip
%
%%%%%%%%%%%%%%%%%%%%%%%%%%%%%%%%%%%%%%%%%%%%%%%%%%%%%%%%%%%%%%%%%%%%%%%%%
% \begin{lemma}...\end{lemma}  does not work properly in the appendix. It gives 
%  ``Lemma Appendix A.1....'' instead of Lemma A.1
%%%%%%%%%%%%%%%%%%%%%%%%%%%%%%%%%%%%%%%%%%%%%%%%%%%%%%%%%%%%%%%%%%%%%%%%%

\noindent\textbf{Lemma A.1} 
\textit{$\tau_N^{m,n}(t)$ can be expressed as follows:}
\begin{eqnarray}
\fl 
\tau^{m,n}_{N}(t)
&=&\prod^{N-1}_{k=1}\left(\frac{q^{n+k-1}b_3}{q^{n+k -1}b_3 -1}\right)^{N-k}\nonumber\\
\fl
&&\times\left|\begin{array}{cccc}
\widetilde\Psi_{m,n}(t) &\widetilde\Psi_{m,n}(qt)& \cdots &\widetilde\Psi_{m,n}(q^{N-1}t)
\end{array}\right|\label{eq:det3}\\
\fl 
&=&\prod^{N-1}_{k=1}\left(\frac{q^{m-k+1}b_1}{q^{m- k+1}b_1 -1}\right)^{N-k}\nonumber\\
\fl
&&\times
\left|\begin{array}{cccc}
\check\Psi_{m,n}(t) &\check\Psi_{m,n+1}(t) & \cdots  &\check\Psi_{m,n+N-1}(t)
\end{array}\right|\label{eq:det4}\\
\fl 
&=&(-1)^{\frac{N(N-1)}{2}}\prod^{N-1}_{k=1}(q^{n+k-1}b_3 )^{N-k}
~\prod^{N-1}_{k=1}\left(\frac{q^{m-k+1}b_1}{q^{m-k+1}b_1-1}\right)^{N-k}\nonumber\\
\fl
&&\times\left|\begin{array}{cccc}
\widehat\Psi_{m,n}(t)&\widehat\Psi_{m,n+1}(q^{-1}t)&\cdots&\widehat\Psi_{m,n+N-1}(q^{-N+1}t)
\end{array}\right|,\label{eq:det5}
\end{eqnarray}
\textit{where the column vectors are given by}
\begin{equation}
\fl
\check\Psi_{m,n}(t)=
\left(\begin{array}{c}
\psi(t,q^mb_1,q^nb_3)\\
\psi(q^{-1}t,q^mb_1,q^{n}b_3)\\
\vdots\\
\psi(q^{-N+1}t,q^mb_1,q^{n}b_3)
\end{array}\right),
\ 
\widehat\Psi_{m,n}(t)=
\left(\begin{array}{c}
\psi(t,q^mb_1,q^nb_3)\\
\psi(qt,q^mb_1,q^{n}b_3)\\
\vdots\\
\psi(q^{N-1}t,q^mb_1,q^{n}b_3)
\end{array}\right),
\end{equation}
\textit{respectively.}\bigskip

\noindent\textbf{Proof.} 
We prove \eref{eq:det3}. 
Using the contiguous relation \eref{eq:re2} to the $N$-th column of the right hand side of
\eref{eq:det2}, we get
\begin{eqnarray*}
\fl
\tau_N^{m,n}(t)&=&
\left|\begin{array}{ccccc}
\widetilde\Psi_{m,n}(t)&\widetilde\Psi_{m,n+1}(t) & \cdots& \widetilde\Psi_{m,n+N-2}(t)
& \frac{q^{n+N-2}b_3 \widetilde\Psi_{m,n+N-2}(qt)-\widetilde\Psi_{m,n+N-2}(t)}{q^{n+N-2}b_3-1}
\end{array}\right|\\
\fl
&=&\frac{q^{n+N-2}b_3} {q^{n+N-2}b_3 -1}
\left|\begin{array}{ccccc}
\widetilde\Psi_{m,n}(t)&\widetilde\Psi_{m,n+1}(t) & \cdots& \widetilde\Psi_{m,n+N-2}(t)
& \widetilde\Psi_{m,n+N-2}(qt)
\end{array}\right|.
\end{eqnarray*}
Applying this procedure up to the second column, we have
\begin{displaymath}
\fl 
\tau_N^{m,n}(t)= \prod^{N-1}_{k=1}\frac{q^{n+k-1}b_3}{q^{n+k-1}b_3 -1}~
\left|
\begin{array}{ccccc}
\widetilde\Psi_{m,n}(t)&\widetilde\Psi_{m,n}(qt) & \cdots & \widetilde\Psi_{m,n+N-3}(qt) &  \widetilde\Psi_{m,n+N-2}(qt)
\end{array}\right|.
\end{displaymath}
Continuing this procedure we obtain
\begin{eqnarray*}
\fl
\tau_N^{m,n}(t)
&=&\left( \prod^{N-1}_{k=1}
\frac{q^{n+k-1}b_3}{q^{n+k-1}b_3 -1} \right)
\times\left( \prod^{N-2}_{k=1}  \frac{q^{n+k-1}b_3}{q^{n+k-1}b_3 -1}\right)
\times  \cdots \times\left(\frac{q^{n}b_3 }{q^{n}b_3 -1}\right)\\
\fl && \hskip40pt \times
\left|\begin{array}{ccccc}
\widetilde\Psi_{m,n}(t)&\widetilde\Psi_{m,n}(qt) & \cdots & \widetilde\Psi_{m,n}(q^{N-2}t) &  
\widetilde\Psi_{m,n}(q^{N-1}t)
\end{array}\right|\\
\fl
&=&\prod^{N-1}_{k=1} \left( \frac{q^{n+k-1}b_3}{q^{n+k-1}b_3 -1}  
\right)^{N-k}\\
\fl
&&\hskip40pt \times
\left|\begin{array}{ccccc}
\widetilde\Psi_{m,n}(t)&\widetilde\Psi_{m,n}(qt) & \cdots & \widetilde\Psi_{m,n}(q^{N-2}t) &  
\widetilde\Psi_{m,n}(q^{N-1}t)
\end{array}\right|,
\end{eqnarray*}
which is \eref{eq:det3}. As to \eref{eq:det4} and \eref{eq:det5} we omit the
details and only describe the method, since one can prove them by the
similar calculations. 
In order to prove \eref{eq:det4} we use the contiguous relation \eref{eq:re1}
on \eref{eq:det1} repeatedly. For \eref{eq:det5} we use \eref{eq:re1} on
\eref{eq:det3} to express $\tau_N^{m,n}(t)$ by the determinant in which 
$t$ is shifted in both horizontal and vertical directions. Finally we
use \eref{eq:re2} on this determinant to derive \eref{eq:det5}. $\square$
\par\medskip

Now the bilinear equations \eref{eq:bl2}-\eref{eq:bl6} can be proved by
the same procedure as that in section \ref{sec:proof}. Therefore
we do not repeat the procedure, but give the list of data which are
necessary for proof of each bilinear equation.  
\par\medskip
\noindent\textbf{\eref{eq:bl2}}
\begin{enumerate}
\item Expression of $\tau_N^{m,n}$: \eref{eq:det2}
\item Difference formula:
\begin{eqnarray}
\fl
&&
\left|\begin{array}{cccc}
\overline{\widetilde\Psi_{m,n}(t)} &\widetilde\Psi_{m-1,n+1}{(qt)} & \cdots & 
\widetilde\Psi_{m-1,n+N-1}{(qt)}
\end{array}\right|\nonumber\\
\fl&&\hskip60pt 
=\frac{ 1}{q^{\frac{(N-1)(2n+N)}{2}}(tb_3)^{N-1}
~{{\displaystyle \prod^{N-1}_{k=0}}}(q^{m-k}b_1  -1)}~\tau^{m,n}_{N}{(t)},\\
\fl
&&
\left|\begin{array}{ccccc}
\widetilde\Psi_{m-1,n+1}(qt) & \overline{\widetilde\Psi_{m,n+1}{(t)}} &\widetilde\Psi_{m-1,n+2}(qt) & 
\cdots & \widetilde\Psi_{m-1,n+N-1}{(qt)}
\end{array}\right|\nonumber\\
\fl
&&\hskip60pt =-\frac{ 1}{\displaystyle q^{\frac{(N-1)(2n+N)}{2}}(tb_3)^{N-1}
{{\displaystyle\prod^{N-1}_{k=0}}} (q^{m-k} b_1 -1) }~\tau^{m,n}_{N}{(t)},
\end{eqnarray}
where
\begin{equation}
\overline{\widetilde\Psi_{m,n}{(t)}}=
\left(
\begin{array}{c} \frac{1}{q^{m}b_1-1}\psi(t,q^mb_1,q^nb_3) \\ \vdots \\\frac{1}{q^{m-N+1} b_1 -1}\psi(t,q^{m-N+1}b_1,q^nb_3) \end{array}
\right). 
\end{equation}
\item Contiguous relation to be used for derivation of difference formula:
\begin{equation}
\psi(t,b_1 ,qb_3)=\psi(t,b_1 ,b_3) + q t b_3 (b_1 -1)~\psi(qt,b_1/q,qb_3).\label{eq:rre1}
\end{equation}
\eref{eq:rre1} can be derived by eliminating $\psi(qt,b_1,qb_3)$ from 
\eref{eq:re1}$_{b_3 \rightarrow qb_3}$ and \eref{eq:re5}.
  \item Pl\"ucker relation:
\begin{eqnarray}
\fl
0&=&
\left|\begin{array}{ccccc}
\widetilde\Psi_{m-1,n}(qt) &\overline{\widetilde\Psi_{m,n}(t)} &
\widetilde\Psi_{m-1,n+1}(qt) &\cdots&\widetilde\Psi_{m-1,n+N-2}{(qt)} \end{array}\right|\nonumber\\
\fl
&&\times  
\left|\begin{array}{ccccc}
\widetilde\Psi_{m-1,n+1}{(qt)}&\widetilde\Psi_{m-1,n+2}(qt) &\cdots &
\widetilde\Psi_{m-1,n+N-1}(qt) & \phi' \end{array}\right|\nonumber\\
\fl
&-&\left|\begin{array}{ccccc}
\widetilde\Psi_{m-1,n}(qt) &\widetilde\Psi_{m-1,n+1}(qt) &\dots & \widetilde\Psi_{m-1,n+N-1}(qt) 
\end{array}\right|\nonumber\\
\fl
&&\times 
\left|\begin{array}{ccccc}
\overline{\widetilde\Psi_{m,n}(t)}&\widetilde\Psi_{m-1,n+1}(qt) &\cdots&
\widetilde\Psi_{m-1,n+N-2}(qt) &\phi' \end{array}\right|\nonumber\\
\fl
&+&
\left|\begin{array}{ccccc}
\widetilde\Psi_{m-1,n}(qt) &\widetilde\Psi_{m-1,n+1}(qt)&\cdots &
\widetilde\Psi_{m-1,n+N-2}(qt)&\phi' \end{array}\right|\nonumber\\
\fl
&&\times 
\left|\begin{array}{ccccc}
\overline{\widetilde\Psi_{m,n}(t)} &\widetilde\Psi_{m-1,n+1}(qt) &\cdots&
\widetilde\Psi_{m-1,n+N-1}(qt) \end{array}\right|,
\end{eqnarray}
where
\begin{equation}
 \phi'=\left(\begin{array}{c}1\\0\\\vdots\\0 \end{array}\right).
\end{equation}
\end{enumerate}

\noindent\textbf{\eref{eq:bl3}}:
\begin{enumerate}
  \item Expression of $\tau_N^{m,n}$: \eref{eq:det2}
  \item Difference formula:
\begin{eqnarray}
\fl
&&
\left|\begin{array}{cccc}
\underline{\widetilde\Psi_{m,n}(t)} &\widetilde\Psi_{m,n+1}{(t/q)} & \cdots & 
\widetilde\Psi_{m,n+N-1}{(t/q)}
\end{array}\right|\nonumber\\
\fl
&&\hskip60pt 
=\frac{t^{N-1} }
{\displaystyle \prod^{N-1}_{k=0} (q^{m-k}tb_1-1)}~\tau^{m,n}_{N}{(t)},\\
\fl
&&
\left|\begin{array}{ccccc}
\widetilde\Psi_{m,n+1}(t/q) & \underline{\widetilde\Psi_{m,n+1}{(t)}} & \widetilde\Psi_{m,n+2}(t/q) & 
\cdots & \widetilde\Psi_{m,n+N-1}{(t/q)}
\end{array}\right|\nonumber\\
\fl
&&\hskip60pt 
=-\frac{t^{N-2} }{\displaystyle \prod^{N-1}_{k=0}
(q^{m-k}tb_1-1)}~\tau^{m,n}_{N}{(t)},
\end{eqnarray}
where
\begin{equation}
\underline{\widetilde\Psi_{m,n}{(t)}}=
\left(
\begin{array}{c} \frac{1}{q^{m}tb_1 -1}\psi(t,q^mb_1,q^nb_1) \\ \vdots \\ 
\frac{1}{q^{m-N+1}t b_1 -1}\psi(t,q^{m-N+1}b_1,q^nb_3) \end{array}
\right).
\end{equation}
\item Contiguous relation to be used for derivation of difference
	formula: \eref{eq:re4}
\item Pl\"ucker relation:
\begin{eqnarray}
\fl
0&=&
\left|\begin{array}{ccccc}
\widetilde\Psi_{m,n}(t/q) &\underline{\widetilde\Psi_{m,n}(t)} &
\widetilde\Psi_{m,n+1}(t/q) &\cdots &\widetilde\Psi_{m,n+N-2}(t/q)
\end{array}\right|\nonumber\\
\fl
&&\times
\left|\begin{array}{ccccc}
\widetilde\Psi_{m,n+1}{(t/q)}&\widetilde\Psi_{m,n+2}(t/q) &\cdots&
\widetilde\Psi_{m,n+N-1}(t/q)&\phi' 
\end{array}\right|\nonumber\\
\fl
&-&
\left|\begin{array}{ccccc}
&\widetilde\Psi_{m,n}(t/q) &\widetilde\Psi_{m,n+1}(t/q) &\cdots&\widetilde\Psi_{m,n+N-1}(t/q) 
\end{array}\right|\nonumber\\
\fl
&&\times 
\left|\begin{array}{ccccc}
\underline{\widetilde\Psi_{m,n}(t)}&\widetilde\Psi_{m,n+1}(t/q) &\cdots&
\widetilde\Psi_{m,n+N-2}(t/q) &\phi' 
\end{array}\right|\nonumber\\
\fl
&+&
\left|\begin{array}{ccccc}
\widetilde\Psi_{m,n}(t/q)&\widetilde\Psi_{m,n+1}(t/q) &\dots &
 \widetilde\Psi_{m,n+N-2}(t/q) & \phi'
\end{array}\right|\nonumber\\
\fl
&&\times
\left|\begin{array}{ccccc}
&\underline{\widetilde\Psi_{m,n}(t)}&\widetilde\Psi_{m,n+1}(t/q)&\cdots& \widetilde\Psi_{m,n+N-1}(t/q) 
\end{array}\right|.
\end{eqnarray}
 \item Derivation of \eref{eq:bl3}: applying the difference formula to
	the Pl\"ucker relation we have
\begin{equation}
\fl
qt ~\tau^{-1,1}_{N}{(t)}~\tau^{0,1}_{N+1}{(qt)}+(1-qtb_1)~\tau^{-1,1}_{N}{(qt)}~\tau^{0,1}_{N+1}{(t)}
-\tau^{-1,2}_{N}{(t)}~\tau^{0,0}_{N+1}{(qt)}=0.\label{eq:re14}
\end{equation}
We obtain \eref{eq:bl3} by eliminating the term $\tau^{-1,1}_{N}{(t)}~\tau^{0,1}_{N+1}{(qt)}$ from
\eref{eq:bl1} and \eref{eq:re14}.
\end{enumerate}

\noindent\textbf{\eref{eq:bl4}:}
\begin{enumerate}
  \item Expression of $\tau_N^{m,n}$: \eref{eq:det1}
  \item Difference formula:
\begin{eqnarray}
\fl
&&
\left|\begin{array}{cccc}
\Psi_{m,n}(t) &\Psi_{m,n-1}{(t)} & \cdots & 
\Psi_{m-N+2,n-1}{(t)}
\end{array}\right|\nonumber\\
\fl
&&\hskip60pt 
=\frac{ {{\displaystyle \prod^{N-2}_{k=0}}} (q^{m-k}b_1  -1)~(1-q^{m-k}tb_1)}
{q^{\frac{(N-1)(2m- N+2)}{2}}b_1^{N-1}}~\tau^{m,n}_{N}{(t)},\\
\fl
&&
\left|\begin{array}{ccccc}
\Psi_{m,n-1}{(t)} &\Psi_{m-1,n}{(t)} &
\Psi_{m-1,n-1}{(t)}&\cdots &\Psi_{m-N+2,n-1}{(t)}
\end{array}\right|\nonumber\\
\fl
&&\hskip60pt 
=\frac{ {{\displaystyle
\prod^{N-2}_{k=1}}} (q^{m-k}b_1 -1)~(1-q^{m-k}t b_1)}{\displaystyle
q^{\frac{(N-1)(2m-N+2)}{2}}b_1^{N-1}}~\tau^{m,n}_{N}{(t)}.
\end{eqnarray}
\item Contiguous relation to be used for derivation of difference
	formula:
\begin{equation}
\psi(t, b_1/q ,b_3 )=\frac{\psi(t , b_1 ,b_3 ) - b_1~\psi(t, b_1,b_3/q)}
{(1- b_1 )(1- t b_1)}.\label{eq:rre3}
\end{equation}
\eref{eq:rre3} can be derived by eliminating $\psi(t/q, b_1 ,b_3)$
from \eref{eq:re1}$_{t \rightarrow t/q}$ and \eref{eq:re4}$_{t\to t/q}$.
\item Pl\"ucker relation:
\begin{eqnarray}
\fl
0&=&
\left|\begin{array}{ccccc}
\Psi_{m+1,n-1}(t)&\Psi_{m,n}(t)&
\Psi_{m,n-1}(t) &\cdots&\Psi_{m-N+3,n-1}{(t)}
\end{array}\right|\nonumber\\
\fl
&&\times 
\left|\begin{array}{ccccc}
\Psi_{m,n-1}{(t)} &\cdots& \Psi_{m-N+3,n-1}(t) &
\Psi_{m-N+2,n-1}(t) &\phi' 
\end{array}\right|\nonumber\\
\fl
&-&
\left|\begin{array}{ccccc}
\Psi_{m+1,n-1}(t) &\Psi_{m,n-1}(t)  &\cdots &
\Psi_{m-N+3,n-1}(t) &\Psi_{m-N+2,n-1}(t)
\end{array}\right|\nonumber\\
\fl
&&\times 
\left|\begin{array}{ccccc}
\Psi_{m,n}(t) &\Psi_{m,n-1}(t) &\cdots&
\Psi_{m-N+3,n-1}(t)&  \phi' 
\end{array}\right|\nonumber\\
\fl
&+&
\left|\begin{array}{ccccc}
\Psi_{m+1,n-1}(t)&\Psi_{m,n-1}(t) &\cdots&
\Psi_{m-N+3,n-1}(t)& \phi' 
\end{array}\right|\nonumber\\
\fl
&&\times 
\left|\begin{array}{ccccc}
\Psi_{m,n}(t) &\Psi_{m,n-1}(t) & \cdots &
\Psi_{m-N+3,n-1}(t)&\Psi_{m-N+2,n-1}(t)
\end{array}\right|.
\end{eqnarray}
\end{enumerate}

\noindent\textbf{\eref{eq:bl5}}:
\begin{enumerate}
  \item Expression of $\tau_N^{m,n}$: \eref{eq:det4}
  \item Difference formula:
\begin{eqnarray}
\fl
&&
\left|\begin{array}{cccc}
{\overline{\check\Psi_{m,n}(t)}} & \check\Psi_{m-1,n+1}(qt) &  \cdots &  
\check\Psi_{m-1,n+N-1}(qt) \end{array}\right|\nonumber\\
\fl
&&\hskip50pt
=\left( q^{n}tb_3 (q^{m}b_1 -1)\right)^{1-N}
\displaystyle{\prod^{N-1}_{k=1}}\left(\frac{q^{m-k+1}b_1}{q^{m-k+1}b_1-1}\right)^{k-N}~\tau_N^{m,n}(t),
\label{eq:d1}\\
\fl
&&\left|\begin{array}{ccccc}
\check\Psi_{m-1,n+1}(qt) & \overline{\check\Psi_{m,n+1}(t)} &   
\check\Psi_{m-1,n+2}(qt)& \cdots &
  \check\Psi_{m-1,n+N-1}(qt)
 \end{array}\right|\nonumber\\
\fl
&&\hskip50pt 
=-\left(q^{n}tb_3 (q^mb_1 -1)\right)^{1-N}
\displaystyle{\prod^{N-1}_{k=1}}\left(\frac{q^{m-k+1}b_1}{q^{m-k+1}b_1-1}\right)^{k-N}~\tau_N^{m,n}(t),
\label{eq:d2}
\end{eqnarray}
where
\begin{equation}
\overline{\check\Psi_{m,n}(q^{l}t)}=
\left(\begin{array}{c} \psi(q^{l}t,q^mb_1,q^nb_3)  \\ q\psi(q^{l-1}t,q^mb_1,q^nb_3) \\ \vdots 
\\q^{N-1}\psi(q^{l-N+1}t,q^mb_1,q^nb_3) \end{array}\right).
\end{equation}
  \item Contiguous relation to be used for derivation of difference
formula:
\begin{equation}
\psi(t , b_1 ,qb_3 )=\psi(t , b_1 ,b_3 ) + qtb_3 ( b_1 -1)~\psi(qt,b_1/q ,qb_3 ).\label{eq:rre4}
\end{equation}
\eref{eq:rre4} can be derived by eliminating $\psi(qt,b_1,qb_3)$ from 
\eref{eq:re1}$_{b_3 \rightarrow qb_3 }$ and \eref{eq:re5}.
  \item Pl\"ucker relation:
\begin{eqnarray}
\fl
0&=&
\left|\begin{array}{ccccc}
\check\Psi_{m-1,n}(qt) &\overline{\check\Psi_{m,n}(t)} &
\check\Psi_{m-1,n+1}(qt) &\dots &  \check\Psi_{m-1,n+N-2}{(qt)}
 \end{array}\right|\nonumber\\
\fl
&&\times
\left|\begin{array}{ccccc}
\check\Psi_{m-1,n+1}{(qt)}&\dots &
\check\Psi_{m-1,n+N-2}(qt)&\check\Psi_{m-1,n+N-1}(qt) &\phi'
 \end{array}\right|\nonumber\\
\fl
&-&
\left|\begin{array}{ccccc}
\check\Psi_{m-1,n}(qt)&\check\Psi_{m-1,n+1}(qt)&\dots&
\check\Psi_{m-1,n+N-2}{(qt)}&\check\Psi_{m-1,n+N-1}{(qt)}
\end{array}\right|\nonumber\\
\fl
&&\times
\left|\begin{array}{ccccc}
\overline{\check\Psi_{m,n}(t)}&\check\Psi_{m-1,n+1}{(qt)}&\dots&
\check\Psi_{m-1,n+N-2}{(qt)} &\phi'
\end{array}\right|\nonumber\\
\fl
&+&
\left|\begin{array}{ccccc}
\check\Psi_{m-1,n}(qt) &\check\Psi_{m-1,n+1}(qt) &
\dots &\check\Psi_{m-1,n+N-2}(qt)& \phi'
\end{array}\right|\nonumber\\
\fl
&&\times
\left|\begin{array}{ccccc}
\overline{\check\Psi_{m,n}(t)}&\check\Psi_{m-1,n+1}(qt) &\dots &
\check\Psi_{m-1,n+N-1}(qt)
\end{array}\right|.
\end{eqnarray}
\end{enumerate}

\noindent\textbf{\eref{eq:bl6}:}
\begin{enumerate}
  \item Expression of $\tau_N^{m,n}$: \eref{eq:det5}
  \item Difference formula:
\begin{eqnarray}
\fl
&&\left|\begin{array}{cccc}
{\overline{\widehat\Psi_{m,n}(t)}} & \widehat\Psi_{m-1,n+1}(t) &  \cdots &  
\widehat\Psi_{m-1,n+N-1}(q^{-N+2}t)
\end{array}\right|\nonumber\\
\fl&&
 =(-1)^{\frac{N(N-1)}{2}}\left(qt (q^{m}b_1 -1)\right)^{1-N}
\displaystyle{\prod^{N-1}_{k=1}}\left(q^{n+k-1}b_3 \right)^{k-N}\nonumber\\
\fl&&
\hskip60pt
\times\displaystyle{\prod^{N-1}_{k=1}}\left(\frac{q^{m-k+1}b_1}{q^{m-k+1}b_1  -1}\right)^{k-N}~\tau_N^{m,n}(t),
\label{eq:dd1}\\
\fl
&&\left|\begin{array}{ccccc}
\widehat\Psi_{m-1,n+1}(t) & \overline{\widehat\Psi_{m,n+1}(t/q)} &   
\widehat\Psi_{m-1,n+2}(t/q)& \cdots &  
\widehat\Psi_{m-1,n+N-1}(q^{-N+2}t)
\end{array}\right|\nonumber\\
\fl
&& =-(-1)^{\frac{N(N-1)}{2}}\Big(qt (q^{m}b_1 -1)\Big)^{1-N} 
\displaystyle{\prod^{N-1}_{k=1}}\left(q^{n+k-1}b_3 \right)^{k-N}\nonumber\\
\fl&&
\hskip60pt
\times\displaystyle{\prod^{N-1}_{k=1}}\left(\frac{q^{m-k+1}b_1}{q^{m-k+1}b_1-1}\right)^{k-N}~\tau_N^{m,n}(t),
\label{eq:dd2}
\end{eqnarray}
where
\begin{equation}
\overline{\widehat\Psi_{m,n}(q^{l}t)}=\left(\begin{array}{c}
\psi(q^{l}t,q^mb_1,q^nb_3)  \\ q^{-1}\psi(q^{l+1}t,q^mb_1,q^nb_3) \\ \vdots 
\\q^{-N+1}\psi(q^{l+N-1}t,q^mb_1,q^nb_3) \end{array}\right).
\end{equation}
\item Contiguous relation to be used for derivation of difference
formula:
\begin{equation}
 \psi(t , b_1 ,b_3) =\psi(qt , b_1 ,b_3/q)+qt(b_1 - 1 )~\psi(qt,  
b_1/q,b_3 ).\label{eq:rre5}
\end{equation}
\eref{eq:rre5} can be derived by eliminating $\psi(qt, b_1 ,b_3)$ from
        \eref{eq:re1} and \eref{eq:re4}.
  \item Pl\"ucker relation:
\begin{eqnarray}
\fl
0&=&
\left|\begin{array}{ccccc}
\widehat\Psi_{m-1,n}(qt)&\overline{\widehat\Psi_{m,n}(t)} &
\widehat\Psi_{m-1,n+1}(t) &\cdots &\widehat\Psi_{m-1,n+N-2}(q^{-N+3}t)
\end{array}\right|\nonumber\\
\fl
&& \times
\left|\begin{array}{ccccc}
\widehat\Psi_{m-1,n+1}{(t)} & \cdots &
\widehat\Psi_{m-1,n+N-2}(q^{-N+3}t)&\widehat\Psi_{m-1,n+N-1}(q^{-N+2}t) &\phi' 
\end{array}\right|\nonumber\\
\fl
&-&
\left|\begin{array}{ccccc}
\widehat\Psi_{m-1,n}(qt)&\widehat\Psi_{m-1,n+1}(t) &
\cdots & \widehat\Psi_{m-1,n+N-2}{(q^{-N+3}t)} &\widehat\Psi_{m-1,n+N-1}(q^{-N+2}t)
\end{array}\right|\nonumber\\
\fl
&& \times
\left|\begin{array}{ccccc}
\overline{\widehat\Psi_{m,n}(t)} &\widehat\Psi_{m-1,n+1}{(t)} &
\dots & \widehat\Psi_{m-1,n+N-2}{(q^{-N+3}t)}& \phi' 
\end{array}\right|\nonumber\\
\fl
&+&
\left|\begin{array}{ccccc}
\widehat\Psi_{m-1,n}(qt) & \widehat\Psi_{m-1,n+1}(t) &\dots &
\widehat\Psi_{m-1,n+N-2}(q^{-N+3}t)&\phi'
\end{array}\right|\nonumber\\
\fl
&& \times 
\left|\begin{array}{ccccc}
\overline{\widehat\Psi_{m,n}(t)}& \widehat\Psi_{m-1,n+1}(t)&
\dots& \widehat\Psi_{m-1,n+N-1}(q^{-N+2}t) 
\end{array}\right|.
\end{eqnarray}
  \item Derivation of \eref{eq:bl6}: applying the above difference
        formula to the Pl\"ucker relation we have
 \begin{equation}
\fl
\tau^{-1,1}_{N}(qt)~\tau^{0,1}_{N+1}(t/q)
+ t (1 - q^{-N} b_1)~\tau^{0,1}_{N}(t)~\tau^{-1,1}_{N+1}(t)
  - \tau^{-1,2}_{N}(t)~\tau^{0,0}_{N+1}(t)=0. \label{eq:bl7}
\end{equation}
We obtain \eref{eq:bl6} by eliminating the term
$\tau^{0,1}_{N}{(t)}~\tau^{-1,1}_{N+1}{(t)}$ from \eref{eq:bl4} and \eref{eq:bl7}.
\end{enumerate}
This completes the proof of Proposition \ref{prop:bl}. $\square$


\section*{References}
\begin{thebibliography}{99}
\bibitem{Adler-Moerbeke} Adler M and van Moerbeke P 2003
Recursion relations for unitary integrals, combinatorics and the Toeplitz lattice
{\it Comm. Math. Phys.}{\bf 237} 397--440.
%
\bibitem{Borodin-Boyarchenko} Borodin A and Boyarchenko D 2003 
Distribution of the first particle in discrete orthogonal polynomial ensembles
{\it Comm. Math. Phys.} {\bf 234} 287--338.
%
\bibitem{Borodin:dP} Borodin A 2003 Discrete gap probabilities and discrete Painlev\'e equations
{\it Duke Math. J.}{\bf 117} 489--542.
%
\bibitem{Chen-Feigin} Chen Y and Feigin M V 2006 Painlev\'e IV and degenerate Gaussian unitary 
ensembles {\it J. Phys. A: Math. Gen.} {\bf 39} 12381-12393.
%
\bibitem{Forrester:growth_and_P} Forrester P J 2003
Growth models, random matrices and {P}ainlev\'e transcendents
{\it Nonlinearity} {\bf 16} R27--R49.
%
\bibitem{Forrester-Witte:biorthogonal} Forrester P J. and Witte N S 2006
Bi-orthogonal polynomials on the unit circle, regular semi-classical
weights and integrable systems {\it Constr. Approx.} {\bf 24} 201--237.
%
\bibitem{Forrester-Witte:P6tau} Forrester P J and Witte N S 2005
Discrete Painlev\'e equations for a class of $P_{\rm VI}$
	$\tau$-functions given as ${\rm U}(N)$ averages
{\it Nonlinearity} {\bf 18} 2061--2088.
%
\bibitem{Forrester-Witte:tau_to_random} Forrester P J and Witte N S 2004
Application of the $\tau$-function theory of Painlev\'e equations to
random matrices: ${\rm P}_{\rm VI}$, the JUE, CyUE, cJUE and scaled limits
{\it Nagoya Math. J.} {\bf 174}  29--114.
%
\bibitem{Forrester-Witte:dP_and_orthogonal} Forrester P J and Witte N S 2004
Discrete Painlev\'e equations, orthogonal polynomials on the unit
	circle, and $N$-recurrences for averages over $U(N)$---${\rm P}_{{\rm III}'}$
	and ${\rm P}_{\rm V} \tau$-functions
{\it Int. Math. Res. Not.} {\bf 2004} 160--183. 
%
\bibitem{Forrester-Witte:discrete} Forrester P J and Witte N S 2003
Discrete Painlev\'e equations and random matrix averages.  {\it	Nonlinearity}  {\bf 16} 1919--1944.
%
\bibitem{Gasper-Rahman} Gasper G and Rahman M 2004 Basic hypergeometric
	series 2nd ed. {\it Encyclopedia of Mathematics and its
	Applications Vol 96} (Cambridge: Cambridge University Press).
%
\bibitem{HKW:qp2} Hamamoto T, Kajiwara K and Witte N S 2006 Hypergeometric
	solutions to the $q$-Painlev\'e equation of type
	$(A_1+A_1^\prime)^{(1)}$ {\it Int. Math. Res. Not.} {\bf 2006}
	Article ID 84169.
%
\bibitem{K:qp3-2} Kajiwara K 2003 On a $q$-difference
	Painlev\'e III equation: II. rational solutions {\it J. Nonlinear Math. Phys. } {\bf 10} 282-303. 
%
\bibitem{KK:qp3-1} Kajiwara K and Kimura K 2003 On a $q$-difference
	Painlev\'e III equation: I. derivation, symmetry and Riccati
	type solutions {\it J. Nonlinear Math. Phys. } {\bf 10} 86-202.
%
\bibitem{KMNOY:hyper1} Kajiwara K, Masuda T, Noumi M, Ohta Y and Yamada
	Y 2004 Hypergeometric solutions to the $q$-Painlev\'e equations
	{\it Int. Math. Res. Not.} {\bf 2004} 2497-2521.
%
\bibitem{KMNOY:hyper2} Kajiwara K, Masuda T, Noumi M, Ohta Y and Yamada
	Y 2005 Construction of hypergeometric solutions to the $q$-Painlev\'e equations
	{\it Int. Math. Res. Not.} {\bf 2005} 1439-1463.
%
\bibitem{KNY:qp4} Kajiwara K, Noumi M and Yamada Y 2001 A study on the
	fourth $q$-Painlev\'e equation {\it J. Phys. A: Math. Gen.} {\bf
	34} 8563-8581.
%
\bibitem{KOS:dP3} Kajiwara K, Ohta Y and Satsuma J 1995 
Casorati determinant solutions for the discrete Painlev\'e III equation
{\it J. Math. Phys.} {\bf 36} 4162-4174.
%
\bibitem{KOSGR:dP2} Kajiwara K, Ohta Y, Satsuma J, Grammaticos B and
	Ramani A 1994 Casorati determinant solutions for the discrete Painlev\'e--II equation
{\it J. Phys. A: Math. Gen.} {\bf 27} 915--922.
%
\bibitem{KYO:dP2} Kajiwara K, Yamamoto K and Ohta Y 1997 Rational
	solutions for the discrete Painlev\'e II equation
  {\it Phys. Lett. A} {\bf 232} 189-199.
%
\bibitem{KTGR:asymmetric} Kruskal M D, Tamizhmani K M, Grammaticos B and Ramani A 2000 
Asymmetric discrete Painlev\'e equations {\it Reg. Chaot. Dyn.} {\bf 5} 273-281.
%
\bibitem{Masuda:private} Masuda T {\it private communication}.
%
\bibitem{MSY} Murata Y, Sakai H and Yoneda J 2002 Riccati solutions of
	discrete Painlev\'e equation with Weyl group symmetry of type
	$E_8^{(1)}$ {\it J. Math. Phys.} {\bf 44} 1396-1414.
%
\bibitem{NSKGR:alt-dP2} Nijhoff F, Satsuma J, Kajiwara K, Grammaticos B
	and Ramani A 1996 A study of the alternate discrete Painlev\'e
	II equation  {\it Inverse Problems} {\bf 12} 697-716.
%
\bibitem{OKS:dP1} Ohta Y, Kajiwara K and Satsuma J 1996 Bilinear
	structure and exact solutions of the discrete Painlev\'e I
	equation {\it  Symmetries and integrability of difference
	equations (CRM Proc. Lect. Notes Vol 9)} ed D Levi, L Vinet and
	P Winternitz (Providence: AMS) 265-268.
%
\bibitem{OHTI:dKP} Ohta Y, Hirota R, Tsujimoto S and Imai T 1993
	Casorati and discrete Gram type determinant representation of
	solutions to the discrete KP hierarchy {\it J. Phys. Soc. Jpn.}
	{\bf 62} 1872-1886.
%
\bibitem{OKMS:RT} Ohta Y, Kajiwara K, Matsukidaira J and Satsuma J 1993 
Casorati determinant solution for the relativistic Toda lattice equation
	{\it J. Math. Phys. }{\bf 34} 5190-5204. 
%
\bibitem{RGO:qp5} Ramani A, Grammaticos B and Ohta Y 2001 The
	 $q$-Painlev\'e V equation and its geometrical description {\it
	J. Phys. A: Math. Gen.} {\bf 34} 2505-2513.
%
\bibitem{RGTT:special_fn} Ramani A, Grammaticos B, Tamizhmani T and
	Tamizhmani K M 2001 Special function solutions of the discrete
	Painlev\'e equations {\it J. Comput. Math. Appl.} {\bf 42}
	603-614.
%
\bibitem{Sakai:qp6_sol} Sakai H 1998 Casorati determinant solutions for
	the $q$-difference sixth Painlev\'e equation  {\it Nonlinearity}  {\bf 11} 823-833. 
%
\bibitem{Sakai} Sakai H 2001 Rational surfaces associated with affine
	root systems and geometry of the Painlev\'e equations {\it
	Commun. Math. Phys.} {\bf 220} 165-229.
%
\bibitem{Tracy-Widom}Tracy C A and Widom H 1999
Random unitary matrices, permutations and Painlev\'e
{\it Comm. Math. Phys.} {\bf 207} 665--685.
%
\end{thebibliography}
\end{document}